\documentclass[published]{nst}

\usepackage{subfigure,dcolumn}
\usepackage[T2A,T1]{fontenc}
\usepackage[russian,english]{babel}

\begin{document}

\title{Simulation Study of Energy Resolution with Changing Pixel Size for Radon Monitor Based on \textit{Topmetal-${II}^-$} TPC}

\doi{10.1007/s41365-018-0532-8}
\author{Mengyao Huang}
\email[Corresponding author, ]{mengyaoh@iastate.edu}
\affiliation{Department of Physics and Astronomy, Iowa State University, Ames, IA, 50010 USA}
\author{Hua Pei}
\affiliation{PLAC, Key Laboratory of Quark $\&$ Lepton Physics (MOE), Central China Normal University, Wuhan, 430079 China}
\author{Xiangming Sun}
\email[Corresponding author, ]{xmsun@phy.ccnu.edu.cn}
\affiliation{PLAC, Key Laboratory of Quark $\&$ Lepton Physics (MOE), Central China Normal University, Wuhan, 430079 China}
\author{Shuguang Zou}
\affiliation{College of Information Science and Engineering, Henan University of Technology, Zhengzhou, 450001 China}

\begin{abstract}
 In this paper, we study how pixel size influences energy resolution for a proposed pixelated detector—a high sensitivity, low cost, and real-time radon monitor based on a \textit{Topmetal-${II}^-$} time projection chamber (TPC). This monitor was designed to improve spatial resolution for detecting radon alpha particles using \textit{Topmetal-${II}^-$} sensors assembled by a 0.35 $\mu$m CMOS integrated circuit process. Owing to concerns that small pixel size might have the side effect of worsening
 energy resolution due to lower signal-to-noise ratio, a Geant4-based simulation was used to investigate the dependence of energy resolution on pixel sizes ranging from 60 $\mu$m to 600 $\mu$m. A non-monotonic trend in this region shows the combined effect of pixel size and threshold on pixels, analyzed by introducing an empirical expression. Pixel noise contributes 50 keV full-width at half-maximum
 energy resolution for 400 $\mu$m pixel size at 1 $\sim$ 4 $\sigma$ threshold that is comparable to
 the energy resolution caused by energy fluctuations in the TPC ionization process ($\sim$ 20 keV). The total energy resolution after combining both factors is estimated to be 54 keV for a pixel size of 400 $\mu$m at 1 $\sim$ 4 $\sigma$ threshold. The analysis presented in this paper would help choosing suitable pixel size for future pixelated detectors.
\end{abstract}


\maketitle

\section{Introduction}\label{sec1}
$^{222}$Rn is a well-known air carcinogen. When radon gas is inhaled, alpha particles emitted by $^{222}$Rn and its progenies will interact with biological tissue in the lungs leading to DNA damage. It is reported by the World Health Organization (WHO) that long-term lung cancer risk rises by about 20$\%$ per 100 Bq/m$^3$ in indoor radon exposure \cite{who}. WHO proposed a reference level of 100 Bq/m$^3$ to minimize the health hazards due to indoor radon exposure, while 200 Bq/m$^3$ is advocated in many countries as an action level \cite{who}. Indoor radon gas can be released naturally from soil adjacent to the foundation, construction materials, and tap water when it is supplied from groundwater in radium-bearing aquifers \cite{radon source}. A long enough 3.8-
day half-life (compared with its short-lived progenies) and the unreactive chemical property of noble gases enables $^{222}$Rn to easily transmit and concentrate in enclosed spaces, and even be inhaled into the human body \cite{indoor radon}. To ensure a safe living environment, it is essential to monitor $^{222}$Rn concentration during and after constructions. It is then necessary to develop an inexpensive, portable, and real-time radon monitor for household and construction supervision.

Radon detectors are categorized according to the time resolutions required for their sampling and analysis including integrating, grab-sampling, and continuous \cite{technique}. Integrating radon detectors (such as SSNTD) only provide monthly or annually averaged radon concentrations, while grab-sampling radon detectors (such as the “Lucas Cells”) take several hours to reach radioactive equilibrium between $^{222}$Rn and its progenies for the desired accuracy. Conversely, continuous radon detectors can be used to obtain $^{222}$Rn concentrations in real time. Commercial products including RAD7 (DURRIDGE, USA), Radon Scout (SARAD GmbH, Germany), and CRM (BARC, India) have sensitivities less than 2 CPH/(Bq$\cdot$m$^{-3}$) \cite{technique}. Recent developments in continuous detectors have focused on obtaining very high sensitivities at relatively low costs via semiconductor schemes \cite{BJT, PIN}. Novel detection approaches such as radioluminescence light have also been proposed but their sensitivities are less than those of commercial products \cite{optical}.

CMOS-based radon detectors that are also promising in cost reduction while retaining high sensitivity have benefited from standard low-cost CMOS foundry processes with high spatial resolution. A bunch of CMOS-based detectors \cite{CMOS on chip, CMOS IC, CMOS pixel, timepix} have been developed and have shown competitive performances. In these designs, $^{222}$Rn or its progenies are collected either passively, or actively using aerosol and electrostatic concentrators, and they are detected by their emitting alpha particles. Time projection chambers (TPCs) have been applied to further improve spatial resolution by making time slices for building up 3D images. Companies include XIA have used this technique to achieve a sensitivity of several alphas/m${^2}$/day for solid materials using pixel sensors of 12 mm on edge \cite{xia}. 

Recently, a pixel sensor called \textit{Topmetal-${II}^-$} has been developed in the Pixel Lab at Central China Normal University \cite{topmetal}. This sensor was assembled by the standard 0.35 $\mu$m CMOS integrated circuit process with pixel noise lower than 15 e$^{-}$. This low noise property enables the sensor to reduce its pixel size to take full advantage of the spatial resolution of micro pixels.

Instead of relying on the assumption of equilibrium between $^{222}$Rn and its progenies, the designed \textit{Topmetal-${II}^-$} TPC radon monitor distinguishes alpha particles from different radioactive elements by combining high-precision 3D imaging with satisfactory energy resolution. Owing to its high sensitivity, it might be able to determine $^{222}$Rn concentrations by only counting alpha particles from $^{222}$Rn to achieve the required precision. As it does not count alpha particles from radioactive progenies of $^{222}$Rn, it is not necessary to wait for hours for radioactive equilibrium to be reached; thus, it could achieve speedy responsiveness. This provides additional robustness under weather conditions where radioactive equilibrium cannot be reached, such as under atmospheric turbulence or relatively high humidity \cite{technique}. It has been reported that precipitation most likely removes $^{218}$Po, $^{214}$Pb, $^{214}$Bi, but not $^{222}$Rn \cite{technique}. However, while spatial resolution can be increased using smaller pixels, it might also affect the energy resolution.

Therefore, we want to use a Geant4-based \cite{geant} simulation method to explore the extent to which changing the pixel size can affect the energy resolution of \textit{Topmetal-${II}^-$} TPC radon detectors. In addition, a non-monotonic trend of energy resolution at small pixel-sized regions is analyzed in detail.

\section{Alpha-detecting \textit{Topmetal-${II}^-$} TPC}\label{sec2}
\begin{figure}[!htb]
	\includegraphics
	[width=0.7\hsize]
	{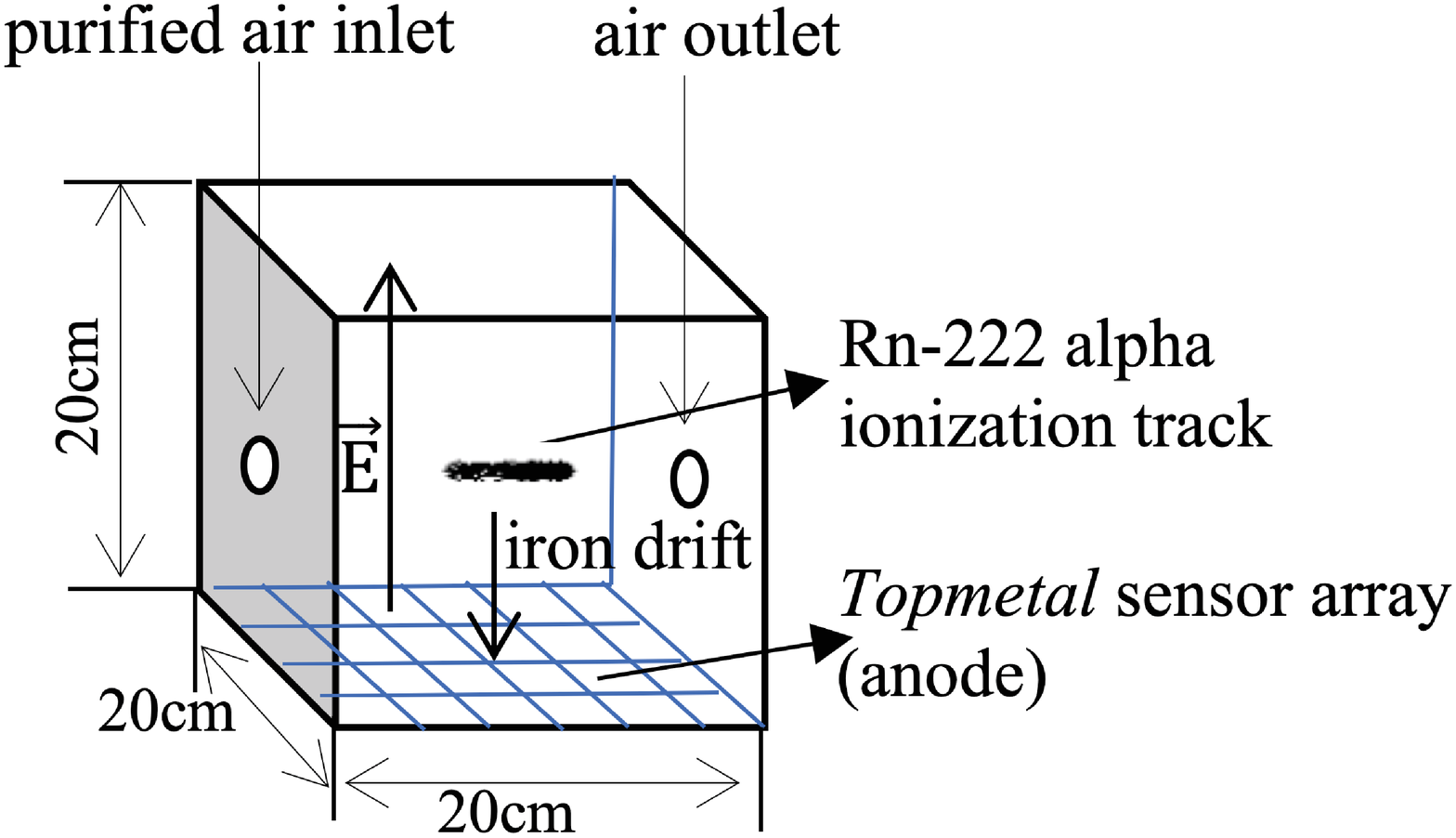}
	\caption{Schematic of radon alpha-detecting \textit{Topmetal-${II}^-$} TPC.}
	\label{fig1}
\end{figure}

Figure~\ref{fig1} is a schematic of radon alpha-detecting \textit{Topmetal-${II}^-$} TPC. The $20\times20\times20$ cm$^3$ cubic volume contains an air sample, with \textit{Topmetal-${II}^-$} pixel sensor arrays placed on the bottom plane. A unique character of \textit{Topmetal-${II}^-$} sensors is that its top material is a metal ($topmetal$) exposed to air, which can serve as an electrode in electric field generation. Another advantage is that the \textit{Topmetal-${II}^-$} sensor is charge sensitive. It can detect both positive and negative charges without requiring free electrons to induce a gas avalanche gain. This is favorable for radon detection, because most of the free electrons created by radon alpha particles will be captured by electronegative molecules in the air during their drift when the drift distance is greater than the mean free path of the electron\cite{ion}. An air supplier is placed on the inlet to provide clean air to the volume. Potentials of -2 kV and 0 V are applied to the top plane and $topmetal$to create an upward uniform electric field of 100 V/cm in between. Within the volume, $^{222}$Rn decays to $^{218}$Po, emitting alpha particles with an energy of 5489 keV. The emitted alpha particles then interact with air molecules to produce ionization electrons, most of which will be attached to electronegative molecules (such as oxygen molecules) during their drift. These negative ions are collected by charge sensitive \textit{Topmetal-${II}^-$} pixel sensor arrays, from which their charge signals are transferred into detectable pulse signals. To simplify our simulation, the more accurate “ion drift” model is replaced by an “electron drift” model. To make this conversion, we need to adjust the sampling rate for the corresponding drift velocity, because ion drift velocity is smaller than electron drift velocity by three to four orders of magnitude \cite{velocity}. Details for this conversion process will be described in Sec.~\ref{sec3}.

\begin{figure}[!htb]
	\includegraphics
	[width=\hsize]
	{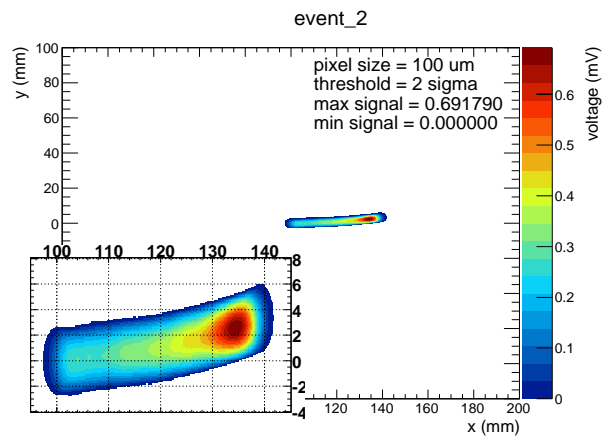}
	\caption{Simulation signal above 2 $\sigma$ threshold (original signal $-$ 2 $\sigma$) on \textit{Topmetal-${II}^-$} pixel sensor array for a single event (event$\_$2) with 100 $\mu$m pixel size. The bottom left panel is an enlarged plot of the signal.}
	\label{fig2}
\end{figure}

Figure~\ref{fig2} is an example of a \textit{Topmetal-${II}^-$} pixel sensor array simulation signal on the bottom charge receiving plane (x-y plane). The size of each pixel is 100 $\mu$m on edge, and a 2 $\sigma$ threshold is applied individually to each pixel. Here, $\sigma$ is defined as the electronic noise on a pixel, which is about 15 e$^{-}$ taken from the test result of \textit{Topmetal-${II}^-$} pixel sensor in \cite{topmetal}. In this paper, the pixel size is characterized by the length of the side of the square pixel. The right color bar shows the signal intensity scale in millivolts. Though the signal in our simulation is in unit of energy (kiloelectronvolt), a charge conversion gain measured in \cite{topmetal} makes it possible to convert energy to voltage. The average minimum ionization energy in the air is 0.0337 keV, i.e., 1 keV energy deposition in the air ionizes about 30 e$^{-}$. Since the \textit{Topmetal-${II}^-$} charge conversion gain is 32.8 e$^{-}$/mV \cite{topmetal}, the conversion between kiloelectronvolt and millivolt is almost 1:1.

In Fig.~\ref{fig2}, an alpha particle of 5489 keV energy is shot from the center parallel to the x-y plane along the x axis. It can be seen that the length of this ionization track is approximately 45 mm. To ensure that the full ionization track leaves the volume, an inner space of 45 mm / 2 = 22.5 mm close to the boundary is eliminated for the rough estimation of sensitivity. This produces a central effective detection volume of about 3.72 L. If a counting efficiency of 100$\%$ can be achieved in this 3.72-L central detection volume, it will produce a maximum sensitivity of $\sim$13.4 CPH/(Bq$\cdot$m$^{-3}$)(3.72 L = $\frac{0.00372}{m^{-3}}\times\frac{3600 s}{h}\times\frac{s^{-1}}{Bq}$ = 13.4 CPH/(Bq$\cdot$m$^{-3}$)). 

This large prototype has been used here to ensure enough data points within a short measurement time. The size of the volume might be reduced to improve portability when we achieve a satisfactory sensitivity. The space between the central detection volume and the boundary can also detect alpha particles, but with less accuracy as most of the particles are cut off by the boundary.

\section{Simulation Process}\label{sec3}
First, a Geant4-based package generates 5489 keV alpha particles inside the volume. Geant4 is a Monte Carlo framework for the simulation of particle passage through matter. To speed up the analysis, the cut-off energy is chosen to be greater than the minimum energy required to produce an electron–ion pair in the air (W value), and this change does not have much effect on the shape and length of the ionization tracks produced by alpha particles. This is because the density of the ionized electrons is sufficiently high, thereby enabling ionized electrons clustering in series to represent the curves of the tracks. After creating a track, each ionization cluster is divided by the W value to obtain the real number of ionized electrons. For simplicity, the ionized electrons are assumed to be spread uniformly inside each cluster.

Assuming that the diffusion of ionized electrons inside the TPC follows a 3D diffusion equation, the expected radius of an electron cluster after diffusion is
\begin{eqnarray}
\label{eq1}
r &=& r_{0}+\sqrt{6Dt},
\end{eqnarray}
where $r_0$ is the initial radius of the cluster, $D$ is electron diffusion coefficient and $t$ is the electron drift time. The electron drift time is calculated as the total drift distance divided by the drift velocity.

The electron diffusion coefficient $D$ and electron drift velocity can be simulated using the Magboltz package \cite{magboltz}. The air parameters are set up as a gas mixture containing 78.08$\%$ of N$_2$, 20.95$\%$ of O$_2$, 0.93$\%$ of Ar and 0.04$\%$ of CO$_2$. Under a vertical electric field of 100 V/cm, room temperature (\SI{20}{\degreeCelsius}) and standard pressure (760.0 Torr), $D$ is found to be 47890.0 mm$^2$/s while the electron drift velocity was 4573000.0 mm/s. The sampling rate of the TPC was set to 457300 Hz to give a spatial resolution of 1 mm in the direction perpendicular to the receiving plane (z-direction). Such spatial resolution setting has been proved to be feasible in the recognition of drifting alpha signals from the experiment for detecting $^{241}$Am by \textit{Topmetal-${II}^-$} TPC in \cite{topmetal}. A sampling rate of 0.6636 ms $\approx$ 1 ms was used in that experiment, with an ion drift velocity of several mm/ms \cite{velocity}; therefore, the corresponding spatial resolution in the z-direction was about 1 mm.

To create signals similar to those expected from real pixel sensors, a 2D grid was coded on the bottom plane, with the size of each grid cell equal to that of a pixel. For simplicity, we assumed that there were no gaps between the sensors and thus the whole $20\times20$ cm$^2$ bottom plane was sensitive to charges. Each electron was collected by the corresponding pixel right under its spatial position after diffusion. Gaussian noise with a mean value of 0 and standard deviation $\sigma$ = 15 e$^{-}$ was added onto each cell to simulate electronic noise in the \textit{Topmetal-${II}^-$} pixel sensor.
It was assumed that the energy of each electron is the same when it reaches the bottom plane as when it was created, neglecting the recombination and decomposition of ions and electrons during the drifting process. The effect of recombination and decomposition may not be negligible for a real detector, but in this paper, we only focus on the dependence of energy resolution on pixel parameters such as size and threshold; therefore, the total energy of the electrons was taken to be the same after the diffusion as before. The fluctuation caused by the ionization process will be counted as an independent factor later in calculating the total energy resolution of the pixelated detector.

The output signal of each pixel at each sampling time is the original charge signal ($O\_Signal_n$) minus the threshold ($T$) placed on each pixel.
\begin{eqnarray}
\label{eq2}
O\_Signal_{n}&=&Noise_{n}+eventEnergy_{n},\\
\label{eq3}
Signal_{n}&=&
\begin{cases}
O\_Signal_{n}-T,&O\_Signal_{n}>T
\cr 
0,&Otherwise
\end{cases},
\end{eqnarray}
where $n$ runs along the corresponding ionization track. A hit is then defined as a nonzero $Signal_{n}$. We assume that a track finding algorithm can be performed to separate tracks with 100$\%$ efficiency. There are two main reasons for this assumption. First, as we force the threshold on each pixel to be > 1 $\sigma$ and less than
the possible maximum signal on the pixel, statistically, this causes $\ge$84.2$\%$ of noises be ruled out in areas that do not receive any external charge. Second, the TPC time resolution provides 3D imaging of an event, and thus, it further suppresses the noise. In particular, the detector will not mix the track of alpha from $^{222}$Rn with alpha from $^{222}$Rn's direct short-lived progeny, $^{218}$Po (half-life 3.05 min), because the time of drifting the farthest negative ions to the charge receiving plane is 200 mm $\div$ several mm/ms < 0.2 s, which is much smaller than 3.05 min.

\begin{figure*}
	\includegraphics
	[width=\hsize]
	{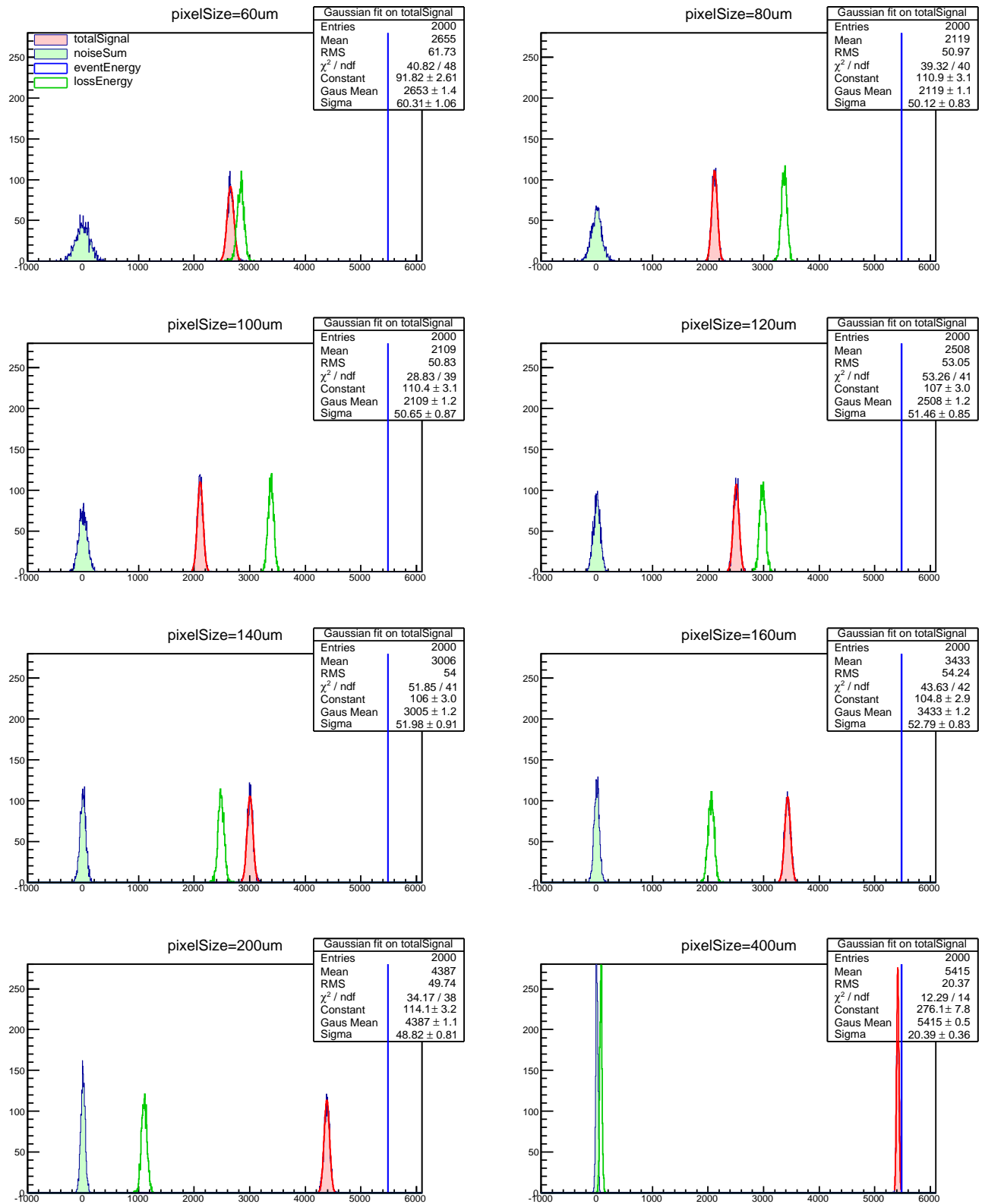}
	\caption{Distribution of total signal ($totalSignal$), total signal without noises and threshold ($eventEnergy$), sum of noises without threshold ($noiseSum$), and total signals that
	deviated from actual energy of an event ($lossEnergy$) for 2000 events, with a 2 $\sigma$ threshold, of pixel size 60 $\mu$m, 80 $\mu$m, 100 $\mu$m, 120 $\mu$m, 140 $\mu$m, 160 $\mu$m, 200 $\mu$m and 400 $\mu$m. Gaussian fitting on the total signal is indicated by the red curve over $totalSignal$. The fitting parameters as well as the statistical parameters are listed in the box at the top right corner of each graph.}
	\label{fig3}
\end{figure*}

To calculate the energy resolution, 2000 radon alpha events are generated from the center of the detector volume with the particle energy for each alpha equal to the $^{222}$Rn-emitting alpha energy (5489 keV). The orientation of the 2000 tracks is all parallel to the bottom pixel plane, because we want to maximize the number of hits so that the fluctuation of the total signal will be most significant to enable a good estimation of the energy resolution. After applying Gaussian noise and energy threshold to each pixel signal at each sampling time, the total signal ($totalSignal$) is calculated by summing up signals of the same event ($Signal_n$). Figure~\ref{fig3} shows that a sample of 2000 events is large enough to give a Gaussian-like shape to the distribution of the $totalSignal$. A least square fit of the Gaussian distribution is performed on the energy spectrum (Fig.~\ref{fig3} red curve), while the full width at half maximum (FWHM) of the Gaussian distribution was used to characterize the energy resolution. Meanwhile, for comparison and quality control, we also calculated the sum of signals without noises and threshold ($eventEnergy$), the sum of noises without threshold ($noiseSum$), and the total signal that deviated from the true energy of an event ($lossEnergy$) for each event. The distribution of $eventEnergy$ is a delta-function mounted at 5489 keV as expected. In all these plots, the $totalSignal$ shifts to a smaller value relative to the $eventEnergy$, and the degree of deviation is calculated as $lossEnergy$. 
\begin{eqnarray}
\label{eq4}
totalSignal&=&\sum_{n}Signal_{n},\\
\label{eq5}
eventEnergy&=&\sum_{n}eventEnergy_{n},\\
\label{eq6}
noiseSum&=&\sum_{n}Noise_{n},\\
\label{eq7}
lossEnergy&=&\sum_{n}(eventEnergy_{n}-Signal_{n}),
\end{eqnarray}
where $n$=1$\cdots$ is the number of hits of one event.

Figure~\ref{fig3} the distributions for the above quantities for different pixel sizes at 2 $\sigma$ threshold, as well as Gaussian fitting on the distribution of $totalSignal$.

\section{Simulation Results}\label{sec4}
\subsection{Energy Resolution versus Pixel Size}
\begin{figure}[!htb]
	\includegraphics
	[width=\hsize]
	{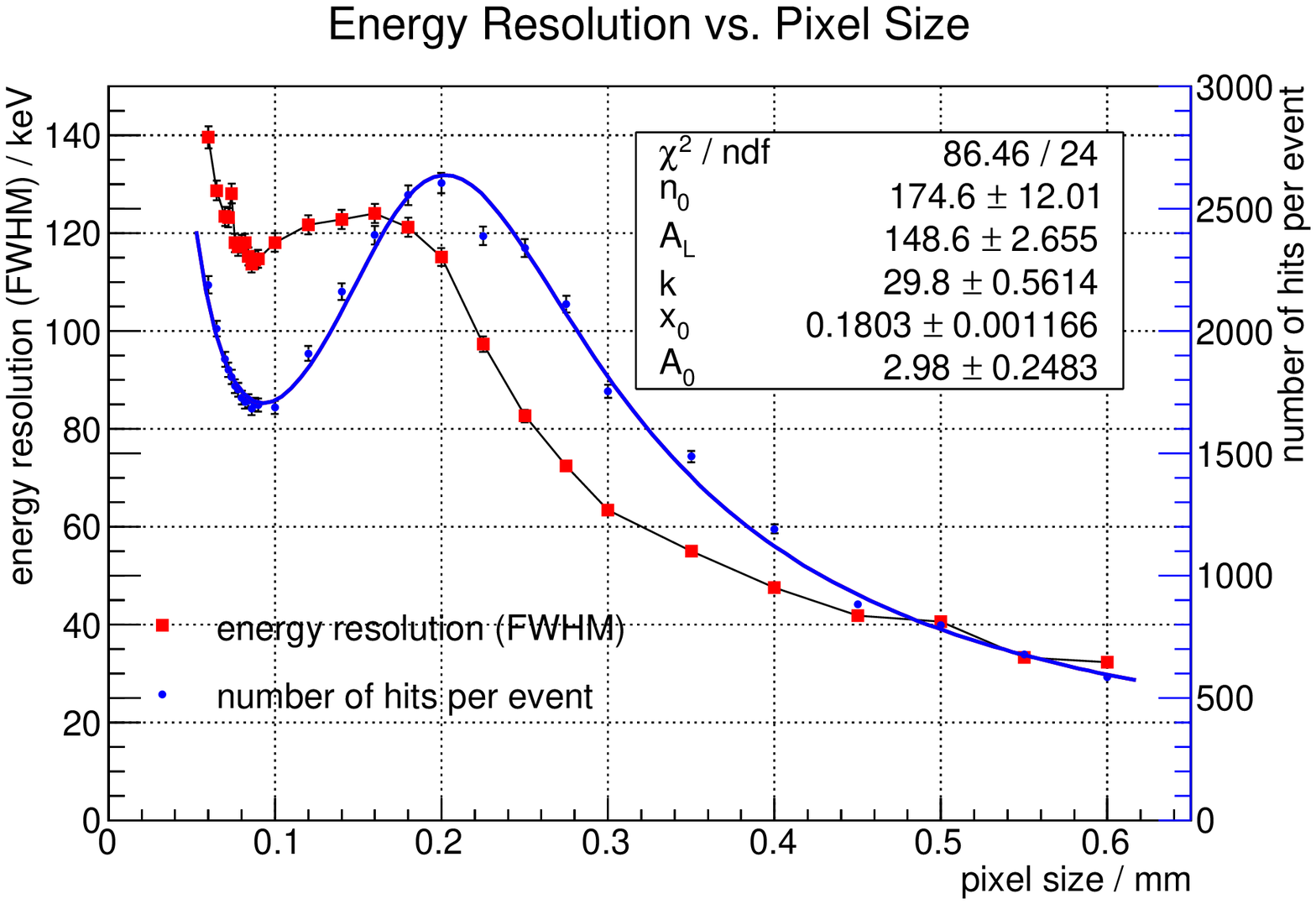}
	\caption{Energy resolution with pixel size (red squares) and number of hits per event with pixel size (blue dots), at 2 $\sigma$ threshold. The blue line represents the fitting line for the number of hits per event.}
	\label{fig4}
	\includegraphics
	[width=\hsize]
	{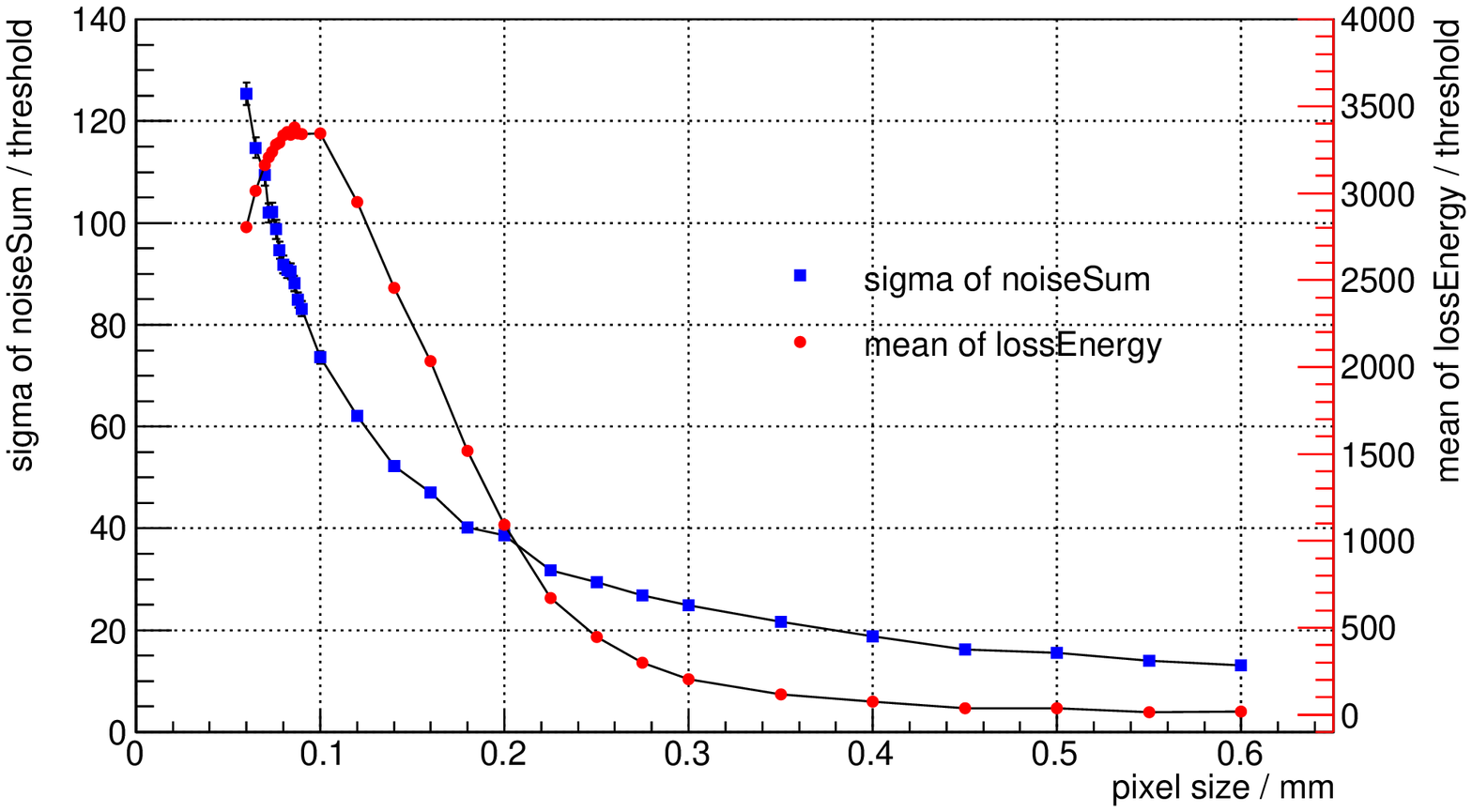}
	\caption{Mean of lossEnergy (red squares) and sigma of noiseSum (blue dots) changing with pixel size, at 2 $\sigma$ threshold.}
	\label{fig5}
\end{figure}

The correlation between the energy resolution and pixel size is shown in Fig.~\ref{fig4} (red squares, left axis scale). The threshold is fixed at a typical value of 2 $\sigma$. Given that $\sigma$ is an intrinsic property of a pixel that depends on the standard foundry process and material, we assumed that $\sigma$ is independent of the pixel size. The energy resolution shows a general decreasing trend in FWHM for very large pixel sizes, in addition to a non-monotonic behavior at small pixel-sized regions due to two counterproductive effects. Energy resolution is expected to be better for larger pixels, because larger pixels receive stronger signals, leading to an increase in the signal-to-noise ratio. Conversely, since a signal is recorded only if it is larger than the threshold, signals generated by larger pixels have higher probabilities of passing the threshold, thus increasing the statistical uncertainty and worsening the energy resolution. However, for pixel sizes that are large enough (>200 $\mu$m in this case), almost all signals are strong enough to pass the threshold, leading to variations due to the second reason to be much less effective. Therefore, the correlation retrieves a monotonic decreasing trend at large pixel sizes.

The variations in energy resolution due to the second reason can be illustrated by plotting the number of hits, i.e., the number of pixels that contain signals larger than the threshold, with the pixel size (blue dots, right axis scale). In addition, to describe this correlation, an empirical expression is introduced for the number of hits per event
\begin{eqnarray}
n&=&n_{0}+\frac{A}{x^{2}},
\end{eqnarray}
where $A$ is the total area that outputs nonzero signals, $x$ represents the pixel size, and $n_{0}$ is a constant due to noise fluctuations. $A$ increases as pixel size increases, because larger pixels receive stronger signals that is favorable to enable the signals to pass the threshold. Thus, $A$ can be expressed as $A=A_{0}+A_{S}$, where $A_{0}$ is a constant at the point where the increase starts. $A_{S}=A_{L}/(1+e^{-k(x-x_{0})} )$ is a raising logistic function with a limit value of $A_{L}$, central point of $x_{0}$ and steepness of $k$. The limit $A_{L}$ is due to the limited total sensing area on the charge-sensing plane for a given event. Figure~\ref{fig4} shows that this expression fits well with the simulation result for the number of hits with a pixel size.

In addition, we carefully analyzed the effect of changing pixel size on $lossEnergy$, because the mean of $lossEnergy$ will be used to calibrate $totalSignal$. As shown in Fig.~\ref{fig5}, the $lossEnergy$ decreases as the pixel size increases. This is because each pixel that has a signal will also contribute a “threshold” value that reduces the total energy. For a given event, the total area that will be hit by the ionization charges is fixed, and therefore, when the pixel size increases, the number of pixels hit by the ionization charges will decrease, causing less “threshold” contribution and $lossEnergy$ to approach 0. This is the main effect when noise does not play a significant role in a signal. However, when the pixel size is less than 100 $\mu$m, single pixel signals are significantly small; therefore, noise will significantly affect them. This is because decreasing pixel sizes implies that more pixels and noises will be added. These noises partially compensate the energy cut-off from thresholds, resulting in a decrease in $lossEnergy$ when pixel size decreases in the small pixel-sized region (<100 $\mu$m in this case). The turning point in Fig.~\ref{fig5} and Fig.~\ref{fig4} shows that for a 2 $\sigma$ threshold, a pixel size around 100 $\sim$ 200 $\mu$m is a boundary about whether noise becomes a dominated factor in deforming the normal trend. In addition, we plot a Gaussian sigma of $noiseSum$, where $noiseSum$ is the sum of original Gaussian noises of all pixels without going through threshold cuts. Since more pixels cause higher statistical fluctuations, the Gaussian sigma for $noiseSum$ is inversely proportional to the pixel size.

\subsection{Energy Resolution versus Threshold on Pixel}
\begin{figure}[!htb]
	\includegraphics
	[width=\hsize]
	{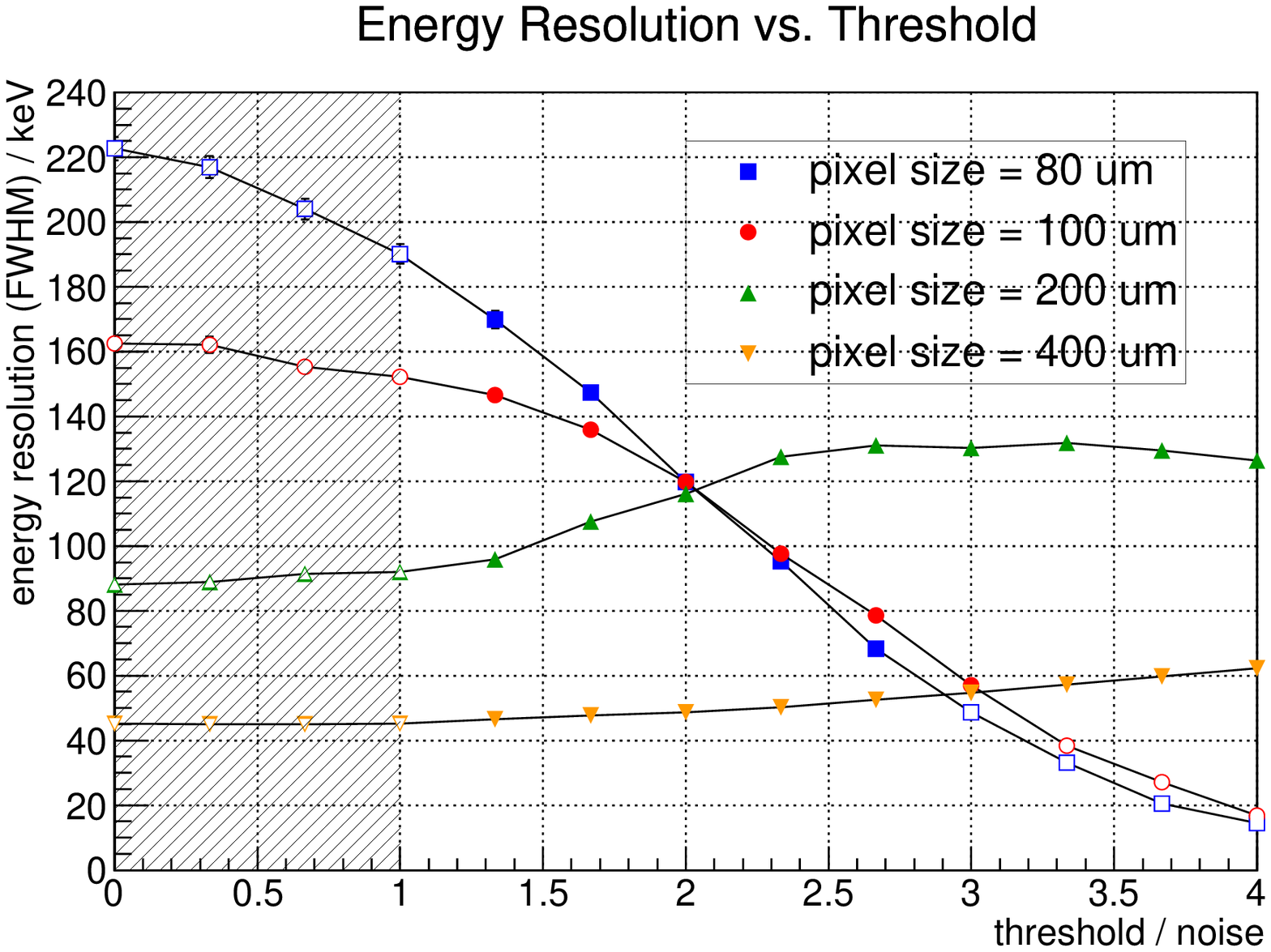}
	\caption{Energy resolution with threshold on pixel.}
	\label{fig6}
	\includegraphics
	[width=\hsize]
	{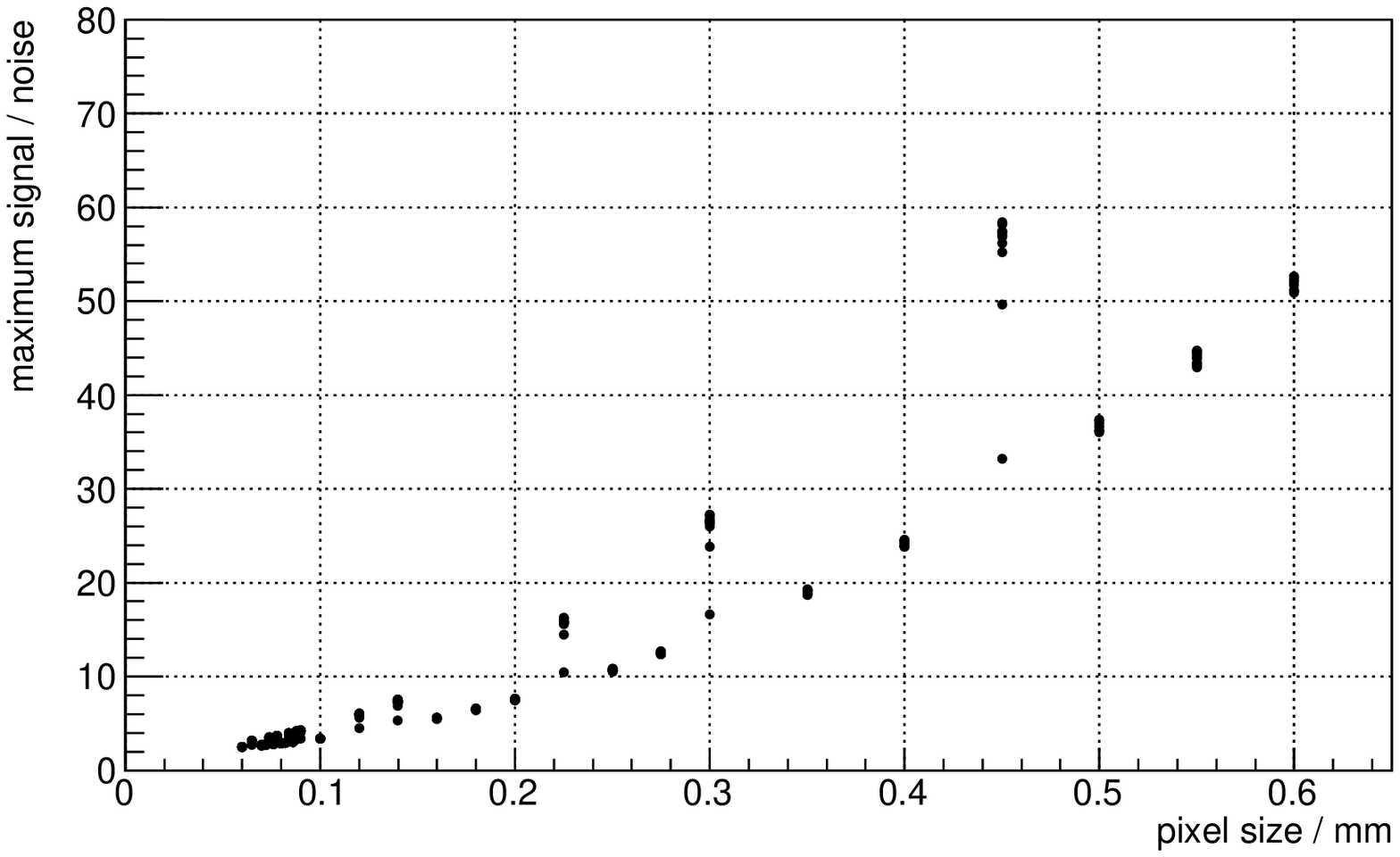}
	\caption{Maximum signal of a pixel before going through threshold cuts (Getting from the signal from 2 $\sigma$ threshold), each pixel size bin has 10 values for event$\_$0 to event$\_$9.}
	\label{fig7}
\end{figure}

The relationship between energy resolution with the threshold on a pixel is shown in Fig.~\ref{fig6}. Four typical pixel sizes of 80 $\mu$m, 100 $\mu$m, 200 $\mu$m and 400 $\mu$m were examined. For pixel sizes of 80 $\mu$m and 100 $\mu$m, the energy resolutions were significantly better at higher thresholds; while for pixel sizes of 200 $\mu$m and 400 $\mu$m, the energy resolutions nearly remained unchanged at all thresholds. This is because while the pixel size is small, noise on the pixel plays a major role in worsening the energy resolution and increasing energy threshold can reduce the noise. Another discovery is that there are several intersecting points between lines, meaning that for the same threshold and the same energy resolution, there exists more than one choice of pixel size. Thresholds lower than 1 $\sigma$ (the shaded area in Fig.~\ref{fig6}) are excluded in our consideration to ensure the effectiveness of the track finding algorithm. Data points whose thresholds are too high also have to be carefully excluded, otherwise the detector has a potential risk of losing a signal.The maximum valid threshold for each pixel size is roughly estimated using the maximum signal recorded at 2 $\sigma$ threshold. For example, in Fig.~\ref{fig2} the maximum signal for 100 $\mu$m with a 2 $\sigma$ threshold (labeled ``max signal'') is observed at about 0.69 keV. Therefore, the maximum signal before going through the 2 $\sigma$ threshold is 0.69 keV + 1.011 keV (energy of 2 $\sigma$ = 2 $\times$ 15 electrons $\times$ 0.0337 keV per electron = 1.011 keV), which is 3.3 $\sigma$. Similarly, for 80 $\mu$m, the maximum signal on a pixel is $\sim$ 2.8 $\sigma$. Since the common threshold for all pixels must be lower than the maximum signal of the pixel, the maximum valid threshold for 80 $\mu$m and 100 $\mu$m is 2.8 $\sigma$ and 3.3 $\sigma$, respectively. Taking this into consideration, the last four data points for 80 $\mu$m and the last three data points for 100 $\mu$m in Fig.~\ref{fig6} should be excluded from our consideration. Using the same method of analysis for 200 $\mu$m and 400 $\mu$m, the maximum pixel signal is greater than 7.5 $\sigma$, so all data points for 200 $\mu$m and 400 $\mu$m in Fig.~\ref{fig6} are valid. Figure~\ref{fig7} shows how the maximum signal-to-noise ratio for a pixel varies with pixel size for 10 events, before any threshold cut is applied. Some maximum signals are kicked to higher values owing to uneven energy distribution on the signal receiving plane and the competitive effect among neighboring pixels.

Among the valid data points (solid points in Fig.~\ref{fig6}), we see that noise on the \textit{Topmetal-${II}^-$} pixel contributes an energy resolution of about 50 keV FWHM for a pixel size of 400 $\mu$m at 1 $\sim$ 4 $\sigma$ threshold. This contribution of energy resolution due to pixel noise is comparable to the energy resolution generated by energy fluctuations in the ionization process ($\sim$ 20 keV). Combining both of them as independent factors, the total energy resolution is $\sim$ 54 keV. 

\section{Compare with Experiment for Alpha Detection of $^{241}\rm Am$}\label{sec5}

$^{241}$Am emits alpha particles with an energy of 5486 keV that is comparable to the radon alpha energy of 5489 keV, and therefore, the simulation result for $^{222}$Rn alpha particles should not deviate too much from the experimental alpha detection for $^{241}$Am. Gao et al. \cite{Am241} show an experiment for detecting alpha particles emitted by $^{241}$Am using \textit{Topmetal-${II}^-$} TPC with a pixel size of $\mu$m. The work shown
in the paper at the current stage did not sum up the energy of each alpha track on pixels, but we can estimate the maximum signal from the detection graph (Figure 14 in \cite{Am241})that is around 4 mV. In Fig.~\ref{fig7}, the maximum signal for a pixel size of about 83 $\mu$m is 2 to 5 $\sigma$. Since 1 $\sigma$ is 15 $\times$ 0.0337 keV = 0.5055 keV $\approx$ 0.5055 mV (the
conversion between kiloelectronvolts and millivolts is almost 1:1 from analysis in Sec.~\ref{sec2}), the maximum signal for radon alpha is 1 to 2.5 mV that is close to the maximum signal from the above experiment for $^{241}$Am alpha. This deviation is largely due to the differences in the experimental setup in \cite{Am241}. In this experiment, alpha particles travel through a small hole in the upper plane; therefore, most of them will go out almost perpendicular to the bottom charge-sensing plane. Even if the longest alpha track is chosen, it still has a very large inclination that will deposit more energy compared to the parallel track in the simulation.

\section{Summary and Outlook}\label{sec6}
We studied how pixel size influences energy resolutions for \textit{Topmetal-${II}^-$} pixelated radon detector when the pixel size is relatively small, using a simulation method based on Geant4. A non-monotonic behavior in energy resolution with pixel size was observed. By fitting the variation of the number of hits with pixel size using an empirical expression that we introduced previously, it can be shown that this phenomenon is due to the combined effect of pixel size and threshold.

The contribution of pixel noise to energy resolution for a pixel size of 400 $\mu$m at 1 $\sim$ 4 $\sigma$ threshold is about 50 keV FWHM that is comparable to the energy resolution caused by energy fluctuations in the ionization process ($\sim$ 20 keV). Treating these two factors that influence energy resolution as independent to each other, the final combining energy resolution is $\sim$ 54 keV. This energy resolution is satisfactory for
distinguishing $^{222}$Rn-alpha particles from alpha particles from other radioactive contaminators in the environment, such as alpha particles of 5305 keV from $^{210}$Po (half-life 138.4 day) and alpha particles of 5686 keV from $^{224}$Ra (half-life 3.7 day), considering both $^{238}$U and $^{232}$Th decay chains. With this good energy resolution, we may also monitor another well-known alpha-emitting health hazard, $^{220}$Rn (half-life 55.6 s) by distinguishing its 6288 keV alpha particle with a 6051/6090 keV alpha particle from $^{212}$Bi (half-life 60.6 min) and 6002 keV alpha particle from $^{218}$Po (half-life 3.05 min). 

Whether it is necessary to use a larger pixel size for better energy resolution also depends on how much a smaller pixel size benefits spatial resolution. In addition, we noticed that this is a simplified model focused on the study of how energy resolution changes with pixel size. Though most of the free electrons are attached to electronegative molecules during their drift, there still might be a small portion of free electrons. The ratio of free electrons to ions might vary with drift distance, adding an additional uncertainty in spatial reconstruction as well as total energy. The boundary conditions for electric field should also be treated properly for a real experiment. More physical processes should be added if energy resolution dependences on other parameters are to be studied.

Another concern with regard to small pixel size is their relatively low signal-to-noise ratios that makes tracking signals on pixels technically difficult before energies from pixels belonging to the same track can be summed up. From Fig.~\ref{fig7}, the maximum signal-to-noise ratio on a pixel ranges from 2 $\sim$ 3 $\sigma$ for a pixel size of about 100 $\mu$m to 50 $\sigma$ for a pixel size of 600 $\mu$m. When the pixel size is larger than 200 $\mu$m, the maximum signal exceeds 7.5 $\sigma$. Furthermore, recent studies on machine learning may also provide a solution to overcome this challenge \cite{ML1,ML2,ML3,ML4}.Characteristics such as straightness and the relative intensity of the energy peak at the end of radon alpha track could be useful patterns in recognizing and tracking the radon alpha signal. In addition, noise performance is improving at the most recent series \textit{Topmetal-${II}a$} \cite{future}; therefore, the overall performance shall be further improved.

\end{document}